\documentclass[12pt]{article}
\usepackage{amssymb}
\usepackage{amsmath,amsfonts,feynmp}
\usepackage{tipa}
\usepackage{stmaryrd}
\usepackage{hyperref}

\makeatletter\@addtoreset{equation}{section}\makeatother

\def\be{\begin{equation}}
\def\ee{\end{equation}}
\def\bea{\begin{eqnarray}}
\def\eea{\end{eqnarray}}

\makeatletter\@addtoreset{equation}{section}\makeatother

\renewcommand{\title}[1]{\vbox{\center\LARGE{#1}}\vspace{5mm}}
\renewcommand{\author}[1]{\vbox{\center#1}\vspace{5mm}}
\newcommand{\address}[1]{\vbox{\center\em#1}}

\begin{document}

\unitlength = .8mm

\begin{titlepage}
\begin{center}
\hfill \\
\hfill \\
\vskip 1cm

\title{ Transport Coefficients from Extremal Gauss-Bonnet Black Holes }

\vskip 0.5cm
 {Rong-Gen Cai$^{a, b, }$\footnote{Email: cairg@itp.ac.cn},
 Yan Liu$^{a, }$\footnote{Email: liuyan@itp.ac.cn}} and {Ya-Wen
Sun$^{a, }$\footnote{Email: sunyw@itp.ac.cn}}

\address{$^{a}$Key Laboratory of Frontiers in Theoretical Physics
\\ Institute of Theoretical Physics, Chinese Academy of Sciences,
\\P.O. Box 2735, Beijing 100190, China}
\address{$^{b}$Department of Physics, Kinki University
\\ Higashi-Osaka, Osaka 577-8502, Japan}
\end{center}

\vskip 1.5cm

\abstract{We calculate the shear viscosity of strongly coupled field
theories dual to Gauss-Bonnet gravity at zero temperature with
nonzero chemical potential. We find that the ratio of the shear
viscosity over the entropy density is $1/4\pi$, which is in
accordance with the zero temperature limit of the ratio at nonzero
temperatures. We also calculate the DC conductivity for this system
at zero temperature and find that the real part of the DC
conductivity vanishes up to a delta function, which is similar to the result in Einstein gravity. We
show that at zero temperature, we can still have the conclusion that
the shear viscosity is fully determined by the effective coupling of
transverse gravitons in a kind of theories that the effective action
of transverse gravitons can be written into a form of minimally
coupled scalars with a deformed effective coupling. }

\vfill

\end{titlepage}


\section{Introduction}
The anti-de Sitter/conformal field theory (AdS/CFT)
\cite{Maldacena:1997re} correspondence has been a quite useful
tool in the study of hydrodynamic properties of strongly coupled
field theories which have dual gravity descriptions. A remarkable
example is the calculation of the ratio of the shear viscosity
over the entropy density, which was found to have a universal
value $1/4\pi$ in a variety of theories described by Einstein
gravity with or without chemical potential
\cite{Policastro:2001yc,{Policastro:2002se},{Policastro:2002tn},{Buchel:2003tz},{Mas:2006dy},{Son:2006em},{Saremi:2006ep},{Maeda:2006by},{Cai:2008in},{Tan:2009yg}}.
With the leading IIB $\alpha'^3$ correction at zero chemical
potential, the value of the ratio has a positive
correction to $1/4\pi$
\cite{{Buchel:2004di},Benincasa:2005qc,{Buchel:2008ac},{Buchel:2008wy},
{Buchel:2008sh},{Myers:2008yi},{Buchel:2008ae}}. This ratio
$1/4\pi$ was conjectured to be a universal lower bound (the KSS
bound) \cite{Kovtun:2003wp,{Kovtun:2004de}} for all materials in
nature. All known materials in nature by now satisfy this bound.
More discussions on the universality and the bound can be found
in~\cite{{Buchel:2004qq},{Cohen:2007qr},{Son:2007vk},{Cherman:2007fj},
{Chen:2007jq},{Son:2007xw},{Fouxon:2008pz},{Dobado:2008ri},{Landsteiner:2007bd}}.

However, in \cite{{KP},Brigante:2007nu,{Brigante:2008gz}}  the
authors calculated the ratio of the shear viscosity over the entropy
density for field theories dual to Gauss-Bonnet gravity using AdS/CFT and found
that the ratio has a negative correction proportional to the
Gauss-Bonnet coupling constant. After taking into the consideration
of causality, a new lower bound $4/25\pi$ was proposed. Later, in
\cite{Brustein:2008cg} it was conjectured that the value of the
shear viscosity is completely determined by the effective coupling
of transverse gravitons at the horizon, which was confirmed in
\cite{Iqbal:2008by} and \cite{Cai:2008ph} using different methods.
The shear viscosity of field theories dual to Gauss-Bonnet gravity
coupled to Maxwell fields was also calculated in \cite{Ge:2008ni}
and in \cite{Cai:2008ph} with $F^4$ corrections, in which the ratio
of the shear viscosity over the entropy density obeys the new lower
bound. Other progress in this aspect can be found in
\cite{{Neupane:2008dc},{Koutsoumbas:2008wy},{Brustein:2008xx},{Buchel:2008vz},{Cai:2009zv},{Banerjee:2009wg},{Ge:2009eh},{Myers:2009ij},
{Cremonini:2009sy},{Banerjee:2009fm},{Ge:2009ac},{Pal:2009qg}}.

However, all the calculations above are in the nonzero temperature
regime, which are not valid at zero temperature.  As argued in a
recent paper \cite{Edalati:2009bi}, taking the small frequency
(hydrodynamic) limit at zero temperature is subtle as the $
\omega\to 0$ and $T\to0$ limits do not, in general, commute.  Thus
one should not just calculate the transport coefficients at nonzero
temperatures and after taking the $T\to 0$ limit state that this
would be the result for the case of $T=0$. Mathematically, this
subtlety arises because the black hole has different singular
structures at the horizon at $T=0$ from at $T>0$, which lead to
different forms of solutions for the hydrodynamic modes. In
\cite{Edalati:2009bi}, transport coefficients of field theories with
nonzero chemical potentials were calculated at zero temperature in
the background of extremal AdS RN black holes using similar methods
as in \cite{Faulkner:2009wj} in the calculations of properties of
non-Fermi fluid\footnote{The original investigation in relating AdS
RN black holes and Fermi surfaces by studying thermodynamic and
transport behaviors was done in \cite{Rey:2009ag}, and later also in
\cite{Lee:2008xf} and \cite{Liu:2009dm,{Cubrovic:2009ye}}. } from AdS/CFT, where the
IR physics plays an important role in deriving the UV physics on the
boundary. They found that at zero temperature, the ratio of the
shear viscosity over the entropy density is $1/4\pi$, which
coincides with the zero temperature limit of the result obtained at
nonzero temperatures. It would be more interesting to study the
shear viscosity for field theories which are dual to Gauss-Bonnet
gravity at zero temperature with nonzero chemical potentials because
at nonzero temperatures the ratio of the shear viscosity over the
entropy density for duals of Gauss-Bonnet gravity has a nontrivial
dependence on the temperature, which is different from the case of
Einstein gravity, where the ratio does not depend on the
temperature. Also it is interesting to investigate whether the
property that the shear viscosity is determined by the effective
coupling of transverse gravitons at the horizon still holds at zero
temperature. These are the motivations of this paper.

In this paper, we calculate the shear viscosity and DC conductivity
for field theories dual to Gauss-Bonnet gravity at zero temperature
with nonzero chemical potentials. We find that the ratio of the
shear viscosity over the entropy density is $1/4\pi$, which does not
depend on the Gauss-Bonnet coupling constant and has the same value
as that for Einstein gravity, which supports the conjecture made
in~\cite{Edalati:2009bi} that the ratio of the shear viscosity over
entropy density is a universal value for extremal black holes. We
also calculate the DC conductivity for this system up to the first
order of the Gauss-Bonnet coupling $\lambda$ and find that the real
part of it vanishes up to a delta function, which is similar to that for Einstein gravity.
Furthermore, we use the same methods to show that at zero
temperature the shear viscosity is also fully determined by the
effective coupling of transverse gravitons in the kind of gravity
theories in which the action of transverse gravitons can be written
into a form of minimally coupled scalars with a deformed effective
coupling which may depend on the radial coordinate. As an
application to the specific system of Gauss-Bonnet gravity coupled
with Maxwell fields, we show that the effective coupling is the same
as Einstein gravity and so the ratio of the shear viscosity over the
entropy density is just $1/4\pi$ at zero temperature, which is
consistent with the result we obtain above.

Our paper is organized as follows. In the next section, we
calculate the shear viscosity and DC conductivity in the
background of an extremal AdS Gauss-Bonnet RN black hole. In
Section 3, we calculate the shear viscosity with a given effective
action of transverse gravitons in the form of a minimally coupled
scalar with an effective coupling. Section 4 is devoted to
conclusions and discussions.

\section{Transport Coefficients from Extremal Gauss-Bonnet RN black holes}
In this section, we calculate the shear viscosity and conductivity
in the background of Extremal Gauss-Bonnet RN black holes in five
dimensions. The action of Gauss-Bonnet gravity coupled with Maxwell
fields in five dimensions is
\cite{Cai,{Cvetic:2001bk},{Astefanesei:2008wz}} \be\label{action}
S=\frac{1}{2\kappa^2}\int
d^5x\sqrt{-g}\Big(R+\frac{12}{\ell^2}+\frac{\lambda}{2}\ell^2(R_{\mu\nu\rho\sigma}R^{\mu\nu\rho\sigma}
-4R_{\mu\nu}R^{\mu\nu}+R^2)-\frac{\kappa^2}{2}
F_{\mu\nu}F^{\mu\nu}\Big),\ee where $\lambda$ is the dimensionless
Gauss-Bonnet coupling and $\kappa^2=8\pi G$. The equations of motion
for this action are
 \bea \label{eom}
  && R_{\mu\nu}-\frac{1}{2}g_{\mu\nu}R-\frac{6}{\ell^2}g_{\mu\nu}+\lambda\ell^2\Big[RR_{\mu\nu}-
2R_{\mu\rho\nu\sigma}R^{\rho\sigma}
+R_{\mu\rho\sigma\tau}R_{\nu}^{~\rho\sigma\tau}-2R_{\mu\rho}R_{\nu}^{~\rho} \nonumber\\
&&~~~~~~~-\frac{1}{4}g_{\mu\nu}(R_{\alpha\beta\rho\sigma}R^{\alpha\beta\rho\sigma}
-4R_{\alpha\beta}R^{\alpha\beta}+R^2)\Big]-\kappa^2\big(F_{\mu\alpha}F_{\nu}^{~\alpha}-\frac{1}{4}g_{\mu\nu}F^2\big)=0,\\
&& \nabla_{\mu}F^{\mu\nu}=0. \eea
The AdS Gauss-Bonnet RN black hole solution with a Ricci flat horizon has the form
 \bea
 && ds^2=-H(r)N^2dt^2+H^{-1}(r)dr^2+\frac{r^2}{\ell^2}(dx^2+dy^2+dz^2),\\
 && A=\mu(1-\frac{r_+^2}{r^2})dt,\eea
 where
 \bea
 && H(r)=\frac{r^2}{2\lambda\ell^2}\Big[1-\sqrt{1-4\lambda\big(1-\frac{m}{r^4}+\frac{q^2}{r^6}\big)}\Big],\nonumber\\
 && N^2=\frac{1}{2}\Big(1+\sqrt{1-4\lambda}\Big),\nonumber\\
 && \mu=\frac{\sqrt{6}qN}{2\kappa\ell r_+^2}\nonumber. \eea
We can rewrite $H(r)$ as
 \be H(r)=\frac{r^2}{2\lambda \ell^2}
 \Big[1-\sqrt{1-4\lambda
\big(1-\frac{r_+^2}{r^2}\big)\big(1-\frac{r_-^2}{r^2}\big)\big(1+\frac{r_+^2+r_-^2}{r^2}\big)}\Big]\nonumber,\ee
where we have $m=r_+^4+\frac{q^2}{r_+^2}$. The thermal properties
of this black hole are \be T=\frac{Nr_+}{2\pi
\ell^2}\Big(2-\frac{q^2}{r_+^6}\Big),
~~s=\frac{2\pi}{\kappa^2}\Big(\frac{r_+}{\ell}\Big)^3,~~\rho=\sqrt{6}\frac{q}{\kappa
\ell^4}, ~~~\epsilon=\frac{3N}{2\kappa^2}\frac{m}{\ell^5}.\ee It
can be easily checked that the first law of thermodynamics
$d\epsilon=Tds+\mu d\rho$ is satisfied.

\subsection{ Extremal Gauss-Bonnet RN black holes}
For $q=\sqrt{2}r_+^3, m=3r_+^4$, we have $T=0$ and the black hole
becomes an extremal one with $r_+=r_-=r_{0}$. In the following of
our calculations it is much more convenient to introduce a
dimensionless parameter $u=r/r_0$. In the new coordinate system
\be f(u)=\frac{1}{2\lambda}\Big[1-\sqrt{1-4\lambda(1-\frac{1}{u^2})^2(1+\frac{2}{u^2})}\Big],\ee
and the solution becomes
 \bea \label{background} ds^2&=&\frac{u^2r_0^2}{\ell^2}\big(-f(u)N^2dt^2+d x^2+dy^2+dz^2\big)+\frac{\ell^2}{u^2}\frac{du^2}{f(u)}, \\
A&=&\mu(1-\frac{1}{u^2})dt, \eea
 where $\mu=\frac{\sqrt{3}r_0N}{\kappa\ell}$.
For this extremal black hole (\ref{background}), the horizon is at
$u=1$ and the boundary is at $u\to \infty$. We introduce
$\alpha=\ell^2/12r_0$ and parameterize the metric by \be
u-1=\lambda_1 \frac{\alpha}{\zeta},~ t=\lambda_1^{-1}\tau.\ee Then
we consider the limit $\lambda_1 \rightarrow 0$ with $\zeta, \tau$
finite, and we can obtain the following near horizon geometry
 \be ds^2=\frac{\alpha r_0}{\zeta^2}\Big[-N^2d\tau^2+d\zeta^2\Big]
 +\frac{r_0}{12\alpha}\Big(dx^2+dy^2+dz^2\Big),~~~~A=\frac{2\mu\alpha}{\zeta}d\tau, \ee
clearly it  has the structure of $\mathrm{AdS}_2\times
\mathrm{R}^3$ with a constant electric field.

\subsection{Shear Viscosity}

In this and the next subsection, we will calculate the shear
viscosity and conductivity of the dual conformal theory to this
extremal Gauss-Bonnet black hole. The shear viscosity can be
calculated from the Kubo formula using the retarded Green's
functions as \be \eta=\lim_{\omega\rightarrow 0}\frac{1}{2\omega
i}\Big(G^A_{xy,xy}(\omega,0)-G^R_{xy,xy}(\omega,0)\Big),\ee where
$\eta$ is the shear viscosity, and the retarded Green's function
is defined by
\begin{equation}
G^R_{\mu\nu,\lambda\rho}(k)=-i\int d^4xe^{-ik\cdot x}\theta (t)
\langle[T_{\mu\nu}(x),T_{\lambda\rho}(0)] \rangle.
\end{equation} The advanced Green's
function can be related to the retarded Green's function of energy
momentum tensor by
$G^A_{\mu\nu,\lambda\rho}(k)=G^R_{\mu\nu,\lambda\rho}(k)^{*}$. In
the frame of AdS/CFT correspondence, one is able to compute the
retarded Green's function of the stress-energy tensor of the dual
field theory by making a small perturbation of the background
metric. Here we work in five dimensions and choose spatial
coordinates so that the momentum of the perturbation points along
the $z$-axis. Then the perturbations can be written as
$h_{\mu\nu}=h_{\mu\nu}(t,z,u)$. In this basis there are three
groups of gravity perturbations in five dimensions, each of which
is decoupled from others: the scalar, vector and tensor
perturbations~\cite{Kovtun:2005ev}. Here we use the simplest one,
the tensor perturbation $h_{xy}$. We use $\phi(t,z,u)$ to denote
this perturbation with one index raised
$\phi(t,z,u)=h_x^{~y}(t,z,u)$. To calculate the Green functions of
the energy momentum tensor at zero temperature we consider the
perturbation $h_x^{~y}$ in the background of (\ref{background}).

We expand
 \be \phi(t,z,u)=\int\frac{dwdp}{(2\pi)^2}e^{-iwt+ipz}\phi(u;k),
 ~~~k=(w,0,0,p),~~~\phi(u;-k)=\phi^*(u;k),\ee
 and by plugging this into the action (\ref{action})
 we can obtain the quadratic order action for $\phi(u)$ as
 \be\label{actionforphi}
  S=-\frac{Nr_0^4}{4\kappa^2\ell^5}\int
  du\frac{dw dp}{(2\pi)^2}u^3 g(u) \Big[\phi^{'2}(u)-
  \frac{\ell^4w^2}{N^2r_0^2f^2(u)u^4}\phi^2(u)\Big]
  \ee up to some total derivatives\footnote{ The explicit form of the total derivative terms need not to be written
out as has been argued in the appendix of  \cite{Cai:2008ph} that they would be canceled after taking into the
consideration of the Gibbons-Hawking boundary term. },  where we have set $p=0$ and
  \be
  g(u)=u^2f(u)\Big[1-2\lambda f(u)-\lambda uf^{'}(u)\Big].
  \ee
 We can easily derive the equation of motion for $\phi(u)$ to be
 \be\label{eomforphi} \phi^{''}(u)+A_1(u) \phi^{'}(u)+B_1(u)\phi(u)=0,\ee
where \bea A_1(u)&=&\Big[\frac{3}{u}+\frac{g^{'}(u)}{g(u)}\Big],\nonumber\\
B_1(u)&=&\frac{144\alpha^2 w^2}{N^2f^2(u)u^4},\eea and the prime $'$
denotes the derivative with respective to $u.$ The analytic solution
of this equation of motion is difficult to obtain, and because we
only need to know the boundary behavior of the solution, we follow
\cite{{Faulkner:2009wj},Edalati:2009bi} to divide the background
spacetime into two regions and match the solution in the overlapping
region. The two regions are defined as

\noindent$\bullet$ near region \be u-1=\frac{\alpha w}{\zeta},
~~~ \epsilon<\zeta<\infty,\ee \noindent $\bullet$ far region \be
u-1>\frac{\alpha w}{\epsilon},\ee and we consider the limit \be
\alpha w\rightarrow 0,~~ ~~\epsilon\rightarrow 0,~~\frac{\alpha
w}{\epsilon}\rightarrow 0,\ee then we can have an overlapping
region

\noindent$\bullet$ over-lapping region
 \be \zeta\rightarrow 0,~~u-1=\frac{\alpha w}{\zeta}\rightarrow 0.\ee

In the near region, we can make a coordinate transformation
$ u-1=\alpha w/\zeta$, and the coefficients $A_1(u)$ and $B_1(u)$ in the
equation of motion (\ref{eomforphi}) become \bea A_1(u)&=&\frac{2}{u-1}
+\tilde{F}_1(u)=\frac{2\zeta}{\alpha w}+F_1(\zeta),\nonumber\\
B_1(u)&=&\frac{\alpha^2w^2}{N^2(u-1)^4}+\frac{\alpha^2 w^2
\tilde{G}_1(u)}{(u-1)^3}=\frac{\zeta^4}{N^2\alpha^2w^2}+\frac{\zeta^3G_1(\zeta)}{\alpha
w},\eea where $\tilde{F}_1(u)=F_1(\zeta)$ and
$\tilde{G}_1(u)=G_1(\zeta)$ are functions regular at $u=1$ which do
not manifestly depend on $\alpha w$. Thus we can expand
$\tilde{F}_1(u)$ and $\tilde{G}_1(u)$ near $u=1$ as
 \bea \tilde{F}_1(u)&=&\tilde{F}_1(1)+\tilde{F}_1^{'}(1)(u-1)+
 \frac{1}{2}\tilde{F}_1^{''}(1)(u-1)^2+\dots,\\
 \tilde{G}_1(u)&=&\tilde{G}_1(1)+\tilde{G}_1^{'}(1)(u-1)+
 \frac{1}{2}\tilde{G}_1^{''}(1)(u-1)^2+\dots.\eea
and expand \bea F_1(\zeta)&=&\tilde{F}_1(1)+\tilde{F}_1^{'}(1)\frac{\alpha w}{\zeta}+\frac{1}{2}\tilde{F}_1^{''}(1)\frac{\alpha^2 w^2}{\zeta^2}+\dots,\\
 G_1(\zeta)&=&\tilde{G}_1(1)+\tilde{G}_1^{'}(1)\frac{\alpha w}{\zeta}
 +\frac{1}{2}\tilde{G}_1^{''}(1)\frac{\alpha^2 w^2}{\zeta^2}+\dots,\eea
 near $\alpha w/\zeta=0.$
Then the equation of motion (\ref{eomforphi}) becomes
 \be\label{phinear} \frac{\partial^2\phi(\zeta)}{\partial\zeta^2}-
 \frac{\alpha w}{\zeta^2}F_1(\zeta)
 \frac{\partial\phi(\zeta)}{\partial \zeta}
 +\Big[\frac{1}{N^2}+\frac{\alpha w}{\zeta}G_1(\zeta)\Big]\phi(\zeta)=0.\ee
 Because we only need to know the low frequency behavior of the
 solution in order to get the shear viscosity, we can expand
 $\phi(\zeta)$ as
 \be\label{nearphiep}
 \phi(\zeta)=\phi^{(0)}(\zeta)+\alpha w\phi^{(1)}(\zeta)+\alpha^2 w^2\phi^{(2)}(\zeta)+\dots .\ee
By plugging (\ref{nearphiep}) into (\ref{phinear}) we can obtain
the following equation to the leading order \be
\frac{\partial^2\phi^{(0)}(\zeta)}{\partial\zeta^2}+\frac{1}{N^2}\phi^{(0)}(\zeta)=0,\ee
with the following general solutions \be\label{sol}
\phi^{(0)}(\zeta)=a_n^{(0)}e^{i\frac{\zeta}{N}}+b_n^{(0)}e^{-i\frac{\zeta}{N}}.\ee
The in-falling boundary condition at the horizon gives
$b_n^{(0)}=0$. We rewrite the solution (\ref{sol}) in the $u$
coordinate and in the matching region we have \be\label{np}
\phi^{(0)}(u)=a_n^{(0)}\Big\{[ 1+\dots]+\frac{i\alpha
w}{(u-1)N}[1+\dots]\Big\}, \ee where the dots represent subleading
terms. We can also calculate the $\phi^{(n)}$ terms order by
order, but knowing the leading order contribution (\ref{np}) is
enough to match the coefficients in the far region.

In the far region, we can also expand $\phi(u)$ as
 \be\label{farphiep}
 \phi(u)=\phi^{(0)}(u)+\alpha w\phi^{(1)}(u)+\alpha^2 w^2\phi^{(2)}(u)+\dots.\ee
By plugging (\ref{farphiep}) into (\ref{eomforphi}) we can obtain
the following equation to the leading order \be
\frac{\partial^2\phi^{(0)}(u)}{\partial^2u}+\Big[\frac{3}{u}+\frac{g^{'}(u)}{g(u)}\Big]\frac{\partial\phi^{(0)}(u)}{\partial
u}=0.\ee The solution to this equation is
\be\label{fphi1}\phi^{(0)}(u)=a_f^{(0)}+b_f^{(0)}\Big[-\frac{1}{12(u-1)}+(-\frac{1}{18}+2\lambda)\ln(u-1)+Y(u)\Big],\ee
where \be
Y^{'}(u)=\frac{1}{u^3g(u)}-\frac{1}{12(u-1)^2}-(-\frac{1}{18}+2\lambda)\frac{1}{u-1}\ee
and the coefficient in (\ref{fphi1}) is chosen in order to make sure
that $Y(u)$ is regular at $u=1$ and  $\phi^{(0)}(u)$ is regular at
$u\to \infty$.

At the boundary $u\to \infty$, the solution (\ref{fphi1}) becomes
\be\label{inphi} \phi^{(0)}(u)|_{u\to
\infty}=a_f^{(0)}+b_f^{(0)}\Big[\lim_{u\to
\infty}\Big((-\frac{1}{18}+2\lambda)\ln(u-1)+Y(u)\Big)-
\frac{1+\sqrt{1-4\lambda}}{8\sqrt{1-4\lambda}}u^{-4}+\dots\Big],\ee
which is the solution we need in order to calculate the shear
viscosity. We need to confirm the coefficients of this solution by
matching the far region solution (\ref{fphi1}) and the near region
solution (\ref{sol}) in the matching region.

In the matching region $u\to 1$, the far region solution
(\ref{fphi1}) becomes \be\label{fp} \phi^{(0)}(u)|_{u\to
1}=\Big[a_f^{(0)}+b_f^{(0)}
Y(1)+\dots\Big]-\frac{b_f^{(0)}}{12(u-1)}\Big[1+\dots\Big].\ee
By
matching (\ref{np}) and (\ref{fp}) we have \be\label{rphi}
a_f^{(0)}=a_n^{(0)}\Big[1+\frac{12i\alpha w}{N}Y(1)\Big], ~~~
b_f^{(0)}=-\frac{12i\alpha w}{N}a_n^{(0)}.\ee Substitute
(\ref{rphi}) into (\ref{inphi}), and we obtain the solution near
the boundary as
\bea \label{resultphi}  \phi(u)|_{u\to \infty}&=&a_n^{(0)}\Big[1+\frac{12i\alpha w}{N}\Big(Y(1)-\lim_{u\to \infty}\big((-\frac{1}{18}+2\lambda)\ln(u-1)+Y(u)\big)\Big)+\dots\Big]\nonumber\\
&&+\frac{3i\alpha w}{2N}a_n^{(0)}\Big[\frac{1+
\sqrt{1-4\lambda}}{\sqrt{1-4\lambda}}+\dots\Big]u^{-4}.\eea

By plugging the boundary solution (\ref{resultphi}) into the
on-shell action for $\phi(u)$ \be
\label{osaforphi}S_{\mathrm{on-shell}}=-\frac{Nr_0^4}{4\kappa^2\ell^5}\int
\frac{dw dp}{(2\pi)^2}u^3 g(u)
\Big(\phi(u)\phi^{'}(u)\Big)\Big|_{u\to\infty},\ee we obtain
\be\mathrm{ Im}G^R_{xy,xy}(w,0)=-\frac{w}{2\kappa^2}
\Big(\frac{r_0}{\ell}\Big)^3[1+\mathcal{O}(w)],\ee and using the
Kubo-formula we have \footnote{At nonzero temperatures, when we take
the $w\to 0$ limit we are comparing $w$ with the temperature $T$ and
in fact taking the $w/T\to 0$ limit. At zero temperature, there is
another dimensionful parameter: the chemical potential. Thus at
$T=0$ when we take the limit of $w\to 0$, we are comparing it with
the chemical potential and this can be viewed as the $w/\mu \to 0$
or $\alpha w\to 0$ limit.} \be\eta=-\lim_{w\to
0}\bigg(\frac{1}{w}\mathrm{Im}G^R_{xy,xy}(w,0)\bigg)=\frac{1}{2\kappa^2}
\Big(\frac{r_0}{\ell}\Big)^3.\ee Note that the entropy density $
s=\frac{2\pi}{\kappa^2}(\frac{r_0}{\ell})^3$, so we have
\be\frac{\eta}{s}=\frac{1}{4\pi}.\ee This result is in accordance
with the $T\rightarrow 0$ behavior of the result obtained in
\cite{Ge:2008ni} for nonzero temperatures though the calculation in
\cite{Ge:2008ni} is not valid for the $T=0$ case. Thus we have the
conclusion that for field theories dual to Gauss-Bonnet gravity at
zero temperature with chemical potential, the ratio of $\eta/s$ is
also $1/4\pi$, which is independent of the Gauss-Bonnet coupling
constant and has the same value as that in Einstein gravity. It will
be shown at the end of Sec.3 explicitly that the ratio of $\eta/s$ we calculate
here is the same as in Einstein theory because the ratio of $\eta/s$
calculated from AdS Gauss-Bonnet black holes depends on the
Gauss-Bonnet coupling constant $\lambda$ only through the product of
$\lambda$ with the $tt$ component of the background metric $g_{tt}$
and the first derivative of $g_{tt}$ on the horizon, and for
extremal black holes $g_{tt}$ and the first derivative of $g_{tt}$
vanish on the horizon. Thus the ratio of $\eta/s$ calculated from
extremal AdS Gauss-Bonnet black holes does not depend on the
parameter $\lambda$ and is of the same value with the ratio from Einstein gravity, which is different from the non-extremal case.

\subsection{DC Conductivity}
In this subsection, we calculate the DC conductivity for the dual field theory in the
background of extremal Gauss-Bonnet RN black holes using a similar
way as in the last subsection. The DC conductivity $ \sigma$ of the dual field theory can be calculated using Kubo formula from the
related Green's function as
\be \sigma=\lim_{\omega\rightarrow 0}\frac{1}{\omega
i}G^R_{x,x}(\omega,0),\ee where
the retarded Green's function $G^R_{x,x}(\omega,0)$ used here
is defined as
\begin{equation}
G^R_{x,x}(k)=-i\int d^4xe^{-ik\cdot x}\theta (t)
\langle[J_x(x),J_x(0)] \rangle \end{equation} and $J_\mu$ is the current operator dual to the bulk gauge field  $A_\mu$.

 To calculate the retarded Green's function defined above we
need to consider small perturbations of the Maxwell field. The
perturbation of the electric fields cannot get decoupled from some
metric perturbations. Thus we consider small metric fluctuations
for components $h_t^{~x}(t,z,u)$ and $h_u^{~x}(t,z,u)$ as well as
the electric field fluctuation $A_x(t,z,u)$. We expand them as
 \bea h_t^{~x}(t,z,u)&=&\int\frac{dwdp}{(2\pi)^2}e^{-iwt+ipz}h_t^{~x}(u;k),\nonumber\\
 h_u^{~x}(t,z,u)&=&\int\frac{dwdp}{(2\pi)^2}e^{-iwt+ipz}h_u^{~x}(u;k),\nonumber\\
 A_x(t,z,u)&=&\int\frac{dwdp}{(2\pi)^2}e^{-iwt+ipz}a_x(u;k),\eea where we choose
 $k=(w,0,0,p), \Phi(u;-k)=\Phi^*(u;k)$ with $\Phi=h_t^{~x}(u;k), h_u^{~x}(u;k), a_x(u;k).$

After choosing a gauge $h_u^{~x}(u)=0$, we get the equations of motion
 for $a_x(u)$ and $h_t^{~x}(u)$ as
  \bea \label{first}
  && \ell^5\kappa w^2 a_x(u)+Nr_0^2uf(u)\Big[\ell\kappa Nu^3a_x^{'}(u)f^{'}(u) \nonumber\\
  &&~~~~~~~~~~~~~~~ +2\sqrt{3}r_0h_t^{~x'}(u)+\ell\kappa Nu^2f(u)\big(3a_x^{'}(u)+ua_x^{''}(u)\big)\Big]=0, \\
\label{second} && 4\sqrt{3}\ell\kappa Na_x(u)+r_0u^5(1-2\lambda
f(u))h_t^{~x'}(u)=0, \eea where the prime $'$ denotes the derivative
with respect to $u.$ From (\ref{second}) we have \be
h_t^{~x'}(u)=-\frac{4\sqrt{3}\ell\kappa
Na_x(u)}{r_0u^5\big(1-2\lambda f(u)\big)},\ee and by plugging this
into (\ref{first}), we obtain the equation of motion for $a_x(u)$
 \be\label{eomforax}
 a_x^{''}(u)+A_2(u) a_x^{'}(u)+B_2(u) a_x(u)=0,
 \ee
where \bea A_2(u)&=&\frac{3}{u}+\frac{f^{'}(u)}{f(u)},
\\
\label{beq}
B_2(u)&=&\frac{144\alpha^2w^2}{N^2u^4f^2(u)}-\frac{24}{u^8f(u)\big(1-2\lambda
f(u)\big)}. \eea We also have to solve this equation by matching the
near region solution with the far region solution.

In the near region we make a coordinate transformation $ u-1=\alpha
w/\zeta$, then we have
 \bea \tilde{A}_2(\zeta)&=&\frac{2}{u-1}+\tilde{F}_2(u)=\frac{2\zeta}{\alpha w}+F_2(\zeta),\nonumber\\
\tilde{B}_2(\zeta)&=&\frac{\alpha^2w^2}{N^2(u-1)^4}+\frac{\alpha^2 w^2 \tilde{G}_2(u)}{(u-1)^3}-\frac{2}{(u-1)^2}+\frac{\tilde{H}_2(u)}{(u-1)}\nonumber\\
&=&\frac{\zeta^4}{N^2\alpha^2w^2}+\frac{\zeta^3G_2(\zeta)}{\alpha
w}-\frac{2\zeta^2}{\alpha^2w^2}+\frac{\zeta{H}_2(\zeta)}{\alpha
w},\eea where $\tilde{F}_2(u)=F_2(\zeta)$,
$\tilde{G}_2(u)=G_2(\zeta)$ and $\tilde{H}_2(u)=H_2(\zeta)$ are
functions regular at $u=1$ and do not depend on $\alpha w$
manifestly. Then we have
 \bea \tilde{F}_2(u)&=&\tilde{F}_2(1)+\tilde{F}_2^{'}(1)(u-1)+\frac{1}{2}\tilde{F}_2^{''}(1)(u-1)^2+\dots,\\
 \tilde{G}_2(u)&=&\tilde{G}_2(1)+\tilde{G}_2^{'}(1)(u-1)+\frac{1}{2}\tilde{G}_2^{''}(1)(u-1)^2+\dots,\\
  \tilde{H}_2(u)&=&\tilde{H}(1)+\tilde{G}_2^{'}(1)(u-1)+\frac{1}{2}\tilde{H}_2^{''}(1)(u-1)^2+\dots, \eea
and in the $\zeta$ coordinate \bea F_2(\zeta)&=&\tilde{F}_2(1)+\tilde{F}_2^{'}(1)\frac{\alpha w}{\zeta}+\frac{1}{2}\tilde{F}_2^{''}(1)\frac{\alpha^2 w^2}{\zeta^2}+\dots,\\
 G_2(\zeta)&=&\tilde{G}_2(1)+\tilde{G}_2^{'}(1)\frac{\alpha w}{\zeta}+\frac{1}{2}\tilde{G}_2^{''}\frac{\alpha^2 w^2}{\zeta^2}+\dots,\\
  H_2(\zeta)&=&\tilde{H}_2(1)+\tilde{H}_2^{'}(1)\frac{\alpha w}{\zeta}+\frac{1}{2}\tilde{H}_2^{''}\frac{\alpha^2 w^2}{\zeta^2}+\dots.\eea

 In the coordinate $\zeta$, the equation of motion (\ref{eomforax}) becomes
 \be\label{axnear} \frac{\partial^2 a_x(\zeta)}{\partial\zeta^2}
 -\frac{\alpha w}{\zeta^2}F_2(\zeta)\frac{\partial a_x(\zeta)}
 {\partial \zeta}+\Big[\frac{1}{N^2}-\frac{2}{\zeta^2}+
 \frac{\alpha w}{\zeta}\bigg(G_2(\zeta)+\frac{H_2(\zeta)}
 {\zeta^2}\bigg)\Big]a_x(\zeta)=0.\ee

We now expand $a_x(\zeta)$ as \be\label{nearaxep}
a_x(\zeta)=a_x^{(0)}(\zeta)+\alpha wa_x^{(1)}(\zeta) +\alpha^2
w^2a_x^{(2)}(\zeta)+\dots,\ee and by plugging (\ref{nearaxep})
into (\ref{axnear}) we obtain the following equation, up to the
leading order, \be
\frac{\partial^2a_x^{(0)}(\zeta)}{\partial\zeta^2}+\Big[\frac{1}{N^2}-\frac{2}{\zeta^2}\Big]a_x^{(0)}(\zeta)=0,\ee
with the following solution \be
a_x^{(0)}(u)=a_{n}^{(0)}\big[1+\frac{iN}{\zeta}\big]e^{i\frac{\zeta}{N}}+b_{n}^{(0)}\big[1-\frac{iN}{\zeta}\big]e^{-i\frac{\zeta}{N}}.\ee
The in-falling boundary condition at the horizon gives
$b_n^{(0)}=0$. In the matching region we can rewrite the near
region solution in the $u$ coordinate as\be\label{nax}
a_x^{(0)}(u)=a_n^{(0)}\Big\{(u-1)[ 1+\dots]+\frac{i\alpha^3
w^3}{3(u-1)^2N^3}[1+\dots]\Big\}. \ee The dots denote subleading
order contributions and we only need the leading order
contributions in the following calculations.

In the far region, we expand $a_x(u)$ as
 \be\label{faraxep}
a_x(u)=a_x^{(0)}(u)+\alpha wa_x^{(1)}(u)+\alpha^2
w^2a_x^{(2)}(u)+\dots.\ee By plugging (\ref{faraxep}) into
(\ref{eomforax}) we obtain the following equation to the leading
order \be \frac{\partial^2a_x^{(0)}(u)}{\partial
u^2}+A_3(u)\frac{\partial a_x^{(0)}(u)}{\partial u}
+B_3(u)a_x^{(0)}(u)=0,
\ee
 where
 \bea
A_3(u)&=&\frac{3}{u}+\frac{f^{'}(u)}{f(u)},\nonumber\\
B_3(u)&=&-\frac{24}{u^8f(u)\big(1-2\lambda f(u)\big)}.\eea It is
difficult to find the exact analytic solution for the above
equation, so we will consider the case of small $\lambda$ and
calculate the first order effect of $\lambda$ in the following
calculations. Now we expand \bea
 &&  a_x^{(0)}(u)=a_{x0}^{(0)}(u)
+\lambda a_{x1}^{(0)}(u)+\mathcal{O}(\lambda^2),\\
 && A_3(u)=\frac{3(2+u^2+u^4)}{u(-1+u^2)(2+u^2)}
 +\lambda \frac{12(-1+u^2)}{u^7}+\mathcal{O}(\lambda^2),\\
 && B_3(u)=-\frac{24}{u^2(-1+u^2)^2(2+u^2)}
  -\lambda \frac{24}{u^8}+\mathcal{O}(\lambda^2).\eea
At the order of $\lambda^{0}$, we have \be
\frac{\partial^2a_{x0}^{(0)}(u)}{\partial
u^2}+\frac{3(2+u^2+u^4)}{u(-1+u^2)(2+u^2)}\frac{\partial
a_{x0}^{(0)}(u)}{\partial u}
-\frac{24}{u^2(-1+u^2)^2(2+u^2)}a_{x0}^{(0)}(u)=0, \ee whose
solution is found to be \be
a_{x0}^{(0)}=a_{f0}^{(0)}\big(1-\frac{1}{u^2}\big)+b_{f0}^{(0)}\Big\{-\frac{19u^4-26u^2+10}{54u^2(u^2-1)^2}+\frac{4(u^2-1)}{81u^2}
\ln\frac{u^2-1}{u^2+2}\Big\}.\ee At the boundary $u\to \infty$, we
have the behavior of the solution as\footnote{Note that there is no
logarithmic divergence term in the boundary solution which was shown
to exist in \cite{Herzog:2009ci} because here we are considering the
leading contribution and the logarithmic divergence term is a
subleading term which comes from the first term in the right hand
side of (\ref{beq}) and does not affect the result.}
\be\label{inax0} a_{x0}^{(0)}(u)|_{u\to
\infty}=a_{f0}^{(0)}-\Big[a_{f0}^{(0)}+\frac{1}{2}b_{f0}^{(0)}\Big]\big(1+\dots\big)u^{-2}\ee
In the matching region $u\to 1$, we have \be\label{fax0}
a_{x0}^{(0)}(u)|_{u\to 1}=
-\frac{b_{f0}^{(0)}}{72(u-1)^2}\Big[1+\mathcal{O}(u-1)\Big]
+(u-1)\Big[2a_{f0}^{(0)}+\frac{b_{f0}^{(0)}}{648}(231
+64\ln\frac{2}{3})+\mathcal{O}(u-1)\Big].\ee On the other hand, at
the order of $\lambda^{1}$, the equation of motion for $a_{x1}^0(u)$
is \bea \frac{\partial^2a_{x1}^{(0)}(u)}{\partial
u^2}+\frac{3(2+u^2+u^4)}{u(-1+u^2)(2+u^2)}\frac{\partial
a_{x1}^{(0)}(u)}{\partial u}
-\frac{24}{u^2(-1+u^2)^2(2+u^2)}a_{x1}^{(0)}(u)&&\nonumber\\
+\frac{12(-1+u^2)}{u^7} \frac{\partial a_{x0}^{(0)}(u)}{\partial
u}-\frac{24}{u^8}a_{x0}^{(0)}(u) =0,~~&& \eea whose solution is
found to be \be
a_{x1}^{(0)}(u)=\frac{1}{2}b_{f0}^{(0)}+a_{f1}^{(0)}\big(1-\frac{1}{u^2}\big)
+b_{f1}^{(0)}\Big\{-\frac{19u^4-26u^2+10}{54u^2(u^2-1)^2}+\frac{4(u^2-1)}{81u^2}
\ln \frac{u^2-1}{u^2+2} \Big \} \ee
 In the limit of $ u\to \infty $, we  have \be\label{inax1}
a_{x1}^{(0)}(u)|_{u\to
\infty}=\big(\frac{1}{2}b_{f0}^{(0)}+a_{f1}^{(0)}\big)
-\Big[a_{f1}^{(0)}+\frac{1}{2}b_{f1}^{(0)}\Big]
\big(1+\dots\big)u^{-2}.\ee
 While in the matching region $u\to 1$,
we have \be\label{fax1}  a_{x1}^{(0)}(u)|_{u\to
1}=-\frac{b_{f1}^{(0)}}{72(u-1)^2}\Big[1
+\mathcal{O}(u-1)\Big]+(u-1)\Big[2a_{f1}^{(0)}
+\frac{b_{f1}^{(0)}}{648}(231+64\ln\frac{2}{3})+\mathcal{O}(u-1)\Big].\ee
Matching (\ref{nax}) to (\ref{fax0}) and (\ref{fax1}) yields
\bea\label{rax} a_{f0}^{(0)}&=&a_n^{(0)}\Big[\frac{1}{2}+\frac{i\alpha^3 w^3}{54}\big(231+64\ln\frac{2}{3}\big)\Big], ~~~ b_{f0}^{(0)}=-24i\alpha^3 w^3a_n^{(0)}, \nonumber\\
a_{f1}^{(0)}&=&a_n^{(0)}\Big[-\frac{i\alpha^3
w^3}{36}\big(231+64\ln\frac{2}{3}\big)\Big], ~~~
b_{f1}^{(0)}=-36i\alpha^3 w^3a_n^{(0)}.\eea Then by substituting
(\ref{rax}) into (\ref{inax0}) and (\ref{inax1}), we have the
boundary behavior of the solution as \bea \label{resultforax} &&
a_x(u)|_{u\to \infty}=a_n^{(0)}\Big[1 +i\alpha^3
w^3\Big(\frac{1}{27}\big(231+64\ln\frac{2}{3}\big)
-\frac{\lambda}{18}\big(663+64\ln\frac{2}{3}\big)\Big)+\dots\Big]\nonumber\\
&&~~ +a_n^{(0)}\Big[-1+i\alpha^3w^3
\Big(\frac{1}{27}\big(417-64\ln\frac{2}{3}\big)
+\frac{\lambda}{18}\big(879+64\ln\frac{2}{3}\big)\Big)
+\dots\Big]u^{-2}.\eea

Substitute this solution (\ref{resultforax}) into the on-shell
action for $a_x$, we obtain  \be
\label{osaforax}S_{\mathrm{on-shell}}=-\frac{Nr_0^2}{2\ell^3}\int
\frac{dw dp}{(2\pi)^2}u^3 f(u)
\Big(a_xa_x^{'}\Big)\Big|_{u\to\infty},\ee and repeat the procedure
in \cite{Brigante:2007nu}, we finally arrive at \be\mathrm{
Im}G^R_{xx}(w,0)\propto w^3[1+\mathcal{O}(w)].\ee As a result, the
DC conductivity behaves in the zero frequency limit as \be
\mathrm{Re}\sigma=-\lim_{w\to
0}\Big(\frac{1}{w}\mathrm{Im}G^R_{xx}(w,0)\Big) \propto \omega^2,\ee
while \be\mathrm{Im} \sigma=-\lim_{w\to
0}\Big(\frac{1}{w}\mathrm{Re}G^R_{xx}(w,0)\Big)\propto
\frac{1}{w}.\ee Because of the Kramers-Kronig relation, the real
part of the DC conductivity calculated above should have a delta
function dependence on $w$ as a result of a pole in the imaginary
part of the conductivity. Thus the real part of the DC conductivity
for duals of Gauss-Bonnet gravity at $T=0$ should vanish up to a
delta function of $w$, which is similar to the case in Einstein
gravity at $T=0$ \cite{Edalati:2009bi} and is in accordance with
\cite{Hartnoll:2007ip} in the limit $T\to 0$.

\section{Universal Properties of $\eta/s$ from Extremal Black Holes}

In Sec.2 we have calculated the shear viscosity and the DC
conductivity for field theories dual to Gauss-Bonnet gravity at
zero temperature with nonzero chemical potential. In this section
we consider a kind of field theories which is dual to gravity
theories in which the effective action of transverse gravitons can
be written into a form of minimally coupled scalars with the
coupling constant deformed to an effective coupling whose value
generally depends on the radial coordinate. In
\cite{{Brustein:2008cg},{Iqbal:2008by},Cai:2008ph}, it was shown
that for this kind of field theories at nonzero temperatures, the
value of the shear viscosity is determined by the value of the
effective coupling at the horizon. The explicit form of the
effective coupling $K_{eff}$ varies for the different gravity
theories, which has been calculated for Einstein and Gauss-Bonnet
gravity coupled to arbitrary matter fields in a very constrained
way in \cite{Cai:2008ph}.

In this section\footnote{In this section we will use the convention
of \cite{Cai:2008ph}, which is  different from the one used in the
last section. We define $v=r_0^2/r^2$ and the horizon is located at
$v=1$ while the boundary is at $v=0$.} we will show that for this
kind of field theories at zero temperature, we can still have the
conclusion that the value of the shear viscosity is proportional to
the effective coupling at the horizon. We  employ the procedure in
\cite{Cai:2008ph}, but for extremal black holes here. We consider
the cases that the perturbation $h_x^{~y}$ can get decoupled from
other modes and first give the assumption of the form of the
effective action of transverse gravitons $h_x^{~y}$ with an
effective coupling $K_{eff}(v)$ and then calculate the shear
viscosity $\eta$ using Kubo formula for transverse gravitons.

We assume that the effective action of transverse gravitons
$\phi(t,v,z)=h_x^{~y}(t,v,z)$ can be written as \be\label{act}
S=\frac{1}{2\kappa^2}\int
dv\frac{dwdp}{(2\pi)^2}\sqrt{-\bar{g}}\bigg(K(v)\phi'\phi'
+w^2K(v)\bar{g}_{vv}\bar{g}^{00}\phi^2-p^2L(v)\phi^2\bigg) \ee up to
some total derivatives, where a prime stands for the derivative with
respect to $v$, $\bar{g}$ denotes the background metric and \be
\phi(t,v,z)=\int
\frac{dwdp}{(2\pi)^2}\phi(v;k)e^{-iwt+ipz},~~k=(w,0,0,p),~~\phi(v;-k)=\phi^*(v;k).
\ee $K(v)$ in this action is related to the effective coupling
$K_{eff}(v)$ by $K_{eff}(v)=K(v)\bar{g}_{vv}$. It was shown in
\cite{Cai:2008ph} that for Einstein gravity with arbitrary minimally
coupled matter fields, $K_{eff}(v)=-1/2$. Thus $K_{eff}(v)$ is a
regular function at the horizon $v=1$ as long as we consider
theories with higher derivative corrections which have contributions
small compared to the Einstein term. Because
$K_{eff}(v)=K(v)\bar{g}_{vv}$ and $K_{eff}(v)$ is regular at the
horizon, we can assume for future use that
$K(v)\sqrt{-\bar{g}}=(1-v)^2M(v)$ for some extremal black hole
backgrounds, where $M(v)$ is a function regular at the horizon.

Because the shear viscosity only involves physics in the zero
momentum limit, $L(v)$ would not affect the value of $\eta$. The
only constraint on $L(v)$ is that it should be regular at the
horizon $v=1$. In fact we can also have an extra term
$w^2N(v)\phi^2$ in the action (\ref{act}), and we assume $N(v)$ is
also a function of $v$,  regular at the horizon $v = 1$. The
addition of such a term will not affect the value of $\eta$.

Because we are working at field theories at zero temperature, the
background metric $\bar{g}$ should be an extremal black hole which
is assumed to be \be
ds^2=-y(v)(1-v)^2dt^2+\frac{1}{h(v)(1-v)^2}dv^2+\frac{r_0^2}{v\ell^2}(d\vec{x}^2),
\ee where $y(v)$ and $h(v)$ take finite values at $v=1$. In this
coordinate system the boundary is at $v=0$ and the horizon is at
$v=1$. This coordinate system can be related to the ordinary one
by a transformation $v=r_0^2/r^2$. The equation of motion for the
transverse gravitons derived from (\ref{act}) is \be\label{eqo}
\phi^{''}(v,k)+A(v)\phi'(v,k)+B(v)\phi(v,k)=0,\ee where \be
A(v)=\frac{(K(v)\sqrt{-\bar{g}})'}{K(v)\sqrt{-\bar{g}}},\ee and
\be B(v)=-\bar{g}_{vv}\bar{g}^{00}w^2+\frac{L(v)}{K(v)}p^2.\ee In
the $p\rightarrow 0$ limit \be B(v)=\frac{w^2}{y(v)h(v)(1-v)^4}.
\ee The behaviors of $A(v)$ and $B(v)$ near the horizon $v=1$ are
quite different from those in the case of non-extremal black
holes. Thus in the near region, the behavior of $\phi(v)$ is
different from that in \cite{Cai:2008ph}. We define the near
region to be $1-v\ll1$ and the far region to be
$1-v\gg\sqrt{\alpha w}$, where $\alpha$ is the same as defined in
last section. In the $\alpha w\ll1$ limit, there exists a matching
region $\sqrt{\alpha w}\ll1-v\ll1$.

In the near region $1-v\ll1$, the equation of motion (\ref{eqo})
becomes\be
\phi^{''}(v)-\frac{2}{1-v}\phi'(v)+\frac{w^2}{y(1)h(1)(1-v)^4}\phi(v)=0.
\ee The solution to this equation has the form
 \be\label{nearu}
\phi(v)=ae^{\frac{iw}{\sqrt{y(1)h(1)}(1-v)}}+be^{-\frac{iw}{\sqrt{y(1)h(1)}(1-v)}},
\ee where $a$ and $b$ are two arbitrary constants. The in-falling
boundary condition sets $b=0$.  On the other hand, in the far
region $1-v\gg \sqrt{\alpha w}$, the equation of motion
(\ref{eqo}) becomes \be \phi^{''}(v)+A(v)\phi'(v)=0.
\ee
  This equation has the solution as  \be \phi(v)=\int
\frac{C}{K(v)\sqrt{-\bar{g}}}dv+C_0,
 \ee
 where $C$ and $C_0$ are two integration constants.

 Note that we have defined $K(v)\sqrt{-\bar{g}}=(1-v)^2M(v)$, where
 $M(v)$ a  function regular at $v=1$. According to the factorization
 rule of rational fractions, $C/\big(K(v)\sqrt{-\bar{g}}\big)$ can be written as
 \be\frac{C}{(1-v)^2M(v)}=\frac{C_1}{(1-v)^2}-\frac{C_2}{1-v}+X(v),\ee
 where $X(v)$ is a function regular at $v=1$. From the regularity of $X(v)$ at
 $v=1$ we can fix the constants $C_1$ and $C_2$ to be
 \be C_1=\frac{C}{M(1)},~~~C_2=\frac{CM'(1)}{M^2(1)}.
 \ee
  Thus $\phi(v)$ can be
 integrated out to be
 \be\label{faru}
 \phi(v)=\frac{C}{M(1)(1-v)}+\frac{CM'(1)}{M^2(1)}\ln{(1-v)}+Z(v),
 \ee where $Z(v)$ is a function regular at $v=1$ and
 \be
 Z'(v)=\frac{C}{(1-v)^2}\bigg(\frac{1}{M(v)}
 -\frac{1}{M(1)}\bigg)+\frac{CM'(1)}{M^2(1)(1-v)}.
 \ee
Then in the matching region $\sqrt{\alpha w}\ll1-v\ll 1$, the near
region solution (\ref{nearu}) can be expanded to the first order of
$\alpha w/(1-v)$ as
 \be\label{nearuu} 1+\frac{iw}{\sqrt{y(1)h(1)}(1-v)},\ee where we have chosen the normalization constant $a$ to be one, while
 the far region solution in the matching region is still of the form
 of (\ref{faru}).
Thus we have \be \frac{C}{M(1)}=\frac{iw}{\sqrt{y(1)h(1)}}, \ee and
$Z(v)=1+iwz(v)$ by matching the leading terms of the near and far
region solutions. Note that the sub-leading terms in the far region
solution (\ref{faru}) need not to appear in the near region solution
as it can be easily checked that the near region solution cannot be
trusted at this order. Thus the solution near the boundary is
\be\label{so}\phi(v)=1+\frac{iw}{\sqrt{y(1)h(1)}(1-v)}+\frac{iw
M'(1)}{M(1)\sqrt{y(1)h(1)}}\ln{(1-v)}+iwz(v),\ee where \be
z'(v)=\frac{M(1)}{\sqrt{y(1)h(1)}(1-v)^2}\bigg(\frac{1}{M(v)}-\frac{1}{M(1)}\bigg)+\frac{M'(1)}
{M(1)\sqrt{y(1)h(1)(1-v)^2}}. \ee Substituting the solution
(\ref{so}) into the on-shell action \be
S_{\mathrm{on-shell}}=\frac{1}{2\kappa^2}\int
\frac{dwdp}{(2\pi)^2}dv\bigg(\sqrt{-\bar{g}}K(v)\phi'\phi\bigg)',\ee
and integrating this action gives \be \label{www}
S_{\mathrm{on-shell}}=\frac{1}{2\kappa^2}\int\frac{dwdp}{(2\pi)^2}
\bigg(\sqrt{-\bar{g}}K(v)\phi'\phi\bigg)\bigg|^{v=1}_{v=0}. \ee As
argued in the appendix of \cite{Cai:2008ph}, the total derivative
terms and the Gibbons-Hawking contribution exactly cancel and
(\ref{www}) is the total contribution. With the help of the boundary
term, we have
 \be
G^R_{xy,xy}(w,0)=-\frac{1}{2\kappa^2}2\sqrt{-\bar{g}}K(v)\phi'^{*}\phi|_{v=0}.
\ee
Substituting the solution (\ref{so}) into the Kubo formula we
finally reach \be \eta=\frac{1}{2\kappa^2}\lim_{w\rightarrow
0}\frac{2\sqrt{-\bar{g}}K(v)
\phi'^{*}\phi|_{v=0}}{iw}=\frac{1}{2\kappa^2}\big(\frac{r_0}{\ell}\big)^3\bigg(-2K_{eff}(v=1)\bigg).
\ee Thus we get to the conclusion that the shear viscosity is
fully determined by the value of the effective coupling
$K_{eff}(v)$ at the horizon. This indicates that the shear
viscosity is totally determined by the IR physics, as we are
considering the $w\rightarrow 0$ limit which encodes only the near
horizon physics.

The form of $K_{eff}(v)$ for some very limited class of gravity
theories have been obtained in \cite{Cai:2008ph}. For Einstein
gravity coupled with matter fields in a constrained way
\cite{Cai:2008ph}, $K_{eff}=-1/2$, so the ratio of $\eta/s$ is
always $1/4\pi$ because in Einstein gravity the entropy density is
proportional to the horizon area. For Gauss-Bonnet gravity coupled
with matter fields in the same constrained way, $K_{eff}(v)$ was
calculated for a specific kind of background metric \be
ds^2=-\frac{g^*(v)r_0^2N^2}{\ell^2v}dt^2+\frac{\ell^2}{4v^2g^*(v)}dv^2+\frac{r_0^2}{v\ell^2}d\vec{x}^2,
\ee
it is
 \be\label{nobody} K_{eff}(v)=-\frac{1}{2}\bigg[1-2\lambda
g^*(v)+2\lambda v g^{*'}(v)\bigg]. \ee
 This form of metric is in
fact the kind of metric which has $g_{tt}g_{rr}=-1$ in the
ordinary $r$ coordinate, which is satisfied by most of the metrics
we are interested in. From the formula (\ref{nobody}) we can see that the value of the shear viscosity has a dependence on $\lambda$ only through the product of $\lambda$ with $g^*(1)$ and $ g^{*'}(1)$ which vanish for extremal black holes. Thus in this specific kind of background
metric we find that $K_{eff}(v)=-1/2$ for extremal black holes,
which means that $\eta/s=1/4\pi$ for the cases that the entropy
density is proportional to the horizon area~\cite{Cai}. This
result is the same as in Einstein gravity as it has no dependence on the Gauss-Bonnet coupling $\lambda$ as we explained above and is in accordance with the result obtained in last section.


\section{Conclusion and Discussion}

In this paper we  calculated the ratio of the shear viscosity over
the entropy density for strongly coupled field theories dual to
Gauss-Bonnet gravity at zero temperature with a nonzero chemical
potential. We found that the ratio is  $1/4\pi$, which does not
depend on the Gauss-Bonnet coupling constant and has the same value
as that for Einstein gravity. We also calculated the DC conductivity
for this system up to the first order of the Gauss-Bonnet coupling
$\lambda$ and found that the real part of it vanishes up to a delta
function, which is similar to the case of Einstein gravity,
again.

We showed that the value of the shear viscosity depends on the
effective coupling of transverse gravitons for a kind of gravity
theories in which the effective action of transverse gravitons can
be written into a form of minimally coupled scalars with a deformed
effective coupling. The effective coupling calculated in Einstein
and Gauss-Bonnet gravity theories with minimally coupled matter
fields shows that the ratio of $\eta/s$ in such theories is always
$1/4\pi$ at zero temperature. The results obtained in this paper
combining with those from \cite{Edalati:2009bi} indicate that those
values of shear viscosity and conductivity are universal for
conformal field theories with gravity duals which have minimally
coupled matter fields at zero temperature\footnote{In the case of
gravity with nonminimally coupled matter fields, at $T=0$ there may
be a violation of the universality of $\eta/s$, which can be seen in
\cite{Myers:2009ij} at the limit $T\to 0$, though the $T\to 0$ limit
may not give the result at $T=0$. }. This is also closely related to
the fact that for extremal black holes there exists an AdS$_2$
factor in the near horizon geometry. In this sense, it would be of
great interest to calculate other transport coefficients for
extremal black holes.

The shear viscosity is fully determined by the near horizon value of
the effective coupling of transverse gravitons. However, we still
have very poor knowledge of the exact form of $K_{eff}$ in various
gravity theories. It would be very interesting to calculate this
quantity in more general gravity theories and in more general kinds
of backgrounds, such as the black hole background in the
non-relativistic version of AdS/CFT.

{\bf Note added:} after this work appeared on arXiv, we notice that
a paper \cite{Paulos:2009yk} appeared on the same day, which
discussed the same topic and reached similar conclusions with ours.
On the next day another paper \cite{Chakrabarti:2009ht} appeared
which has some overlap with our present paper.

\section*{Acknowledgments}

We would like to thank Robert G. Leigh and C. Herzog for useful
correspondences. RGC thanks L. M. Cao and N. Ohta for useful
discussions.  This work was supported partially by grants from
NSFC, China (No. 10535060, No. 10821504 and No. 10975168) and a
grant from MSTC, China (No. 2010CB833004). This work was completed
during RGC's visit to Kinki University, Japan with the support of
JSPS invitation fellowship, the warm hospitality extended to him
is greatly appreciated.


\end{document}